\documentclass[aps,prl,twocolumn,10pt,amsmath,superscriptaddress,amssymb,longbibliography]{revtex4-2}
\usepackage{epsfig}
\usepackage{epstopdf}
\usepackage{dcolumn}
\usepackage{amsbsy}
\usepackage{bm}
\usepackage{color,CJK}
\usepackage{amsmath,amssymb,amsfonts}
\usepackage{appendix}
\usepackage{graphicx}
\usepackage{algorithm}
\usepackage{enumerate}
\usepackage{makeidx}
\usepackage{braket}
\usepackage[usenames,dvipsnames]{xcolor}
\usepackage[driverfallback=dvipdfmx,colorlinks,linkcolor=blue,citecolor=blue,urlcolor=blue]{hyperref}

\begin{document}

\begin{CJK*}{GB}{gbsn}

\title{Chiral Raman coupling for spin-orbit coupling in ultracold atomic gases}

\author{Biao Shan}
\affiliation{State Key Laboratory of Quantum Optics Technologies and Devices, \\  Institute of Opto-electronics, Collaborative Innovation Center of Extreme Optics, Shanxi University, Taiyuan, Shanxi 030006, China}

\author{Lianghui Huang}
\email[Corresponding author email: ]{huanglh06@sxu.edu.cn;}
\affiliation{State Key Laboratory of Quantum Optics Technologies and Devices, \\  Institute of Opto-electronics, Collaborative Innovation Center of Extreme Optics, Shanxi University, Taiyuan, Shanxi 030006, China}

\author{Yuhang Zhao}
\affiliation{State Key Laboratory of Quantum Optics Technologies and Devices, \\  Institute of Opto-electronics, Collaborative Innovation Center of Extreme Optics, Shanxi University, Taiyuan, Shanxi 030006, China}

\author{Guoqi Bian}
\affiliation{State Key Laboratory of Quantum Optics Technologies and Devices, \\  Institute of Opto-electronics, Collaborative Innovation Center of Extreme Optics, Shanxi University, Taiyuan, Shanxi 030006, China}

\author{Pengjun Wang}
\affiliation{State Key Laboratory of Quantum Optics Technologies and Devices, \\  Institute of Opto-electronics, Collaborative Innovation Center of Extreme Optics, Shanxi University, Taiyuan, Shanxi 030006, China}

\author{Wei Han}
\affiliation{State Key Laboratory of Quantum Optics Technologies and Devices, \\  Institute of Opto-electronics, Collaborative Innovation Center of Extreme Optics, Shanxi University, Taiyuan, Shanxi 030006, China}

\author{Jing Zhang}
\email[Corresponding author email: ]{jzhang74@sxu.edu.cn}
\affiliation{State Key Laboratory of Quantum Optics Technologies and Devices, \\  Institute of Opto-electronics, Collaborative Innovation Center of Extreme Optics, Shanxi University, Taiyuan, Shanxi 030006, China}
\affiliation{Hefei National Laboratory, Hefei, China.}

\date{\today }

\begin{abstract}
Spin-orbit coupling (SOC) in ultracold atoms is engineered by light-atom interaction, such as two-photon Raman transitions between two Zeeman spin states.
In this work, we propose and experimentally realize chiral Raman coupling to generate SOC in ultracold atomic gases, which exhibits high quantization axis direction-dependence.
Chiral Raman coupling for SOC is created by chiral light-atom interaction, in which a circularly polarized electromagnetic field generated by two Raman lasers interacts with two Zeeman spin states $\delta m_{F}=\pm 1$ (chiral transition).
We present a simple scheme of chiral one-dimension (1D) Raman coupling by employing two Raman lasers at an intersecting angle 90$^{\circ}$ with the proper polarization configuration.
In this case, Raman coupling for SOC exist in one direction of the magnetic quantization axis and disappears in the opposite direction.
Then we extend this scheme into a chiral 2D optical square Raman lattice configuration to generate the 1D SOC.
There are two orthogonal 1D SOC, which exists in the positive and negative directions of the magnetic quantization axis respectively.
This case is compared with 2D SOC based on the nonchiral 2D optical Raman lattice scheme for studying the topological energy band.
This work broadens the horizon for understanding chiral physics and simulating topological quantum systems.

\end{abstract}

\maketitle
\end{CJK*}

\section{I. Introduction}
Chirality plays a critical role in a wide range of systems, from matter, light, and light-matter interaction~\cite{Kimoon2012CR,Geoffrey2016Science}, which displays a kind of symmetry breaking characterized by lacking mirror-reflection symmetry~\cite{Qiu2023Nature} and has been attracting intense attention in a broad range of scientific areas.
The recent discovery of exotic chiral matters and phenomena involving chiral superconductors~\cite{kallin2016RPP}, chiral skyrmions~\cite{shibata2013NN}, chiral domain walls ~\cite{chen2013NC,emori2013NM,Parkin2013NN}, chiral spintronics~\cite{Parkin2021NRP}, chiral currents~\cite{Bloch2014NP}, chiral electrons~\cite{Dai2015PRX}, nanoscale chiral valley-photon interface~\cite{Kuipers2018Sci}, and chiral quantum optics~\cite{Peter2017Nat}, have aroused widespread interest. In particular, interaction between chiral light and chiral matter leads to chiral light-matter interaction, which triggers exiting research directions and new applications~\cite{Strangi2023AM,Smirnova2019NPO,Park2015PRL}.


Recent experimental realization of spin-orbit coupling (SOC) in ultracold quantum gases by light-matter interaction ~\cite{Spielman2011nat,Wang2012PRL,Martin2012PRL,Huang2016a,Benjamin2016PRX,Shuai2016Sci,Cooper2016PRL,
FallaniPRL2016,Lin2018PRL,shuai2021sci,zhai2012MPB,Congjun2013JPBAMOP,pengjun2014FP} provides a highly controllable platform for the study of topological materials and exotic matter states~\cite{Dalibard2011RMP,Jung2012PRL,Spielman2013,Spielman2014RPP,zhai2015RPP,Meng2016PRL,Ketterle2016PRL,Pu2016PRA,Engels2014nc,Wolfgang2017Nature,Han2018PRL,Huang2018PRA,Chuanwei2019nc,Spielman2021NC}.
In this paper, we explore chiral Raman coupling for SOC in ultracold atomic gases by chiral light-matter interaction, in which a circularly polarized electromagnetic field generated by two Raman lasers interacts with two Zeeman spin states $\delta m_{F}=\pm 1$.
Chiral Raman coupling presents characteristics with high quantization axis direction-dependence.
We employ two Raman lasers at an intersecting angle 90$^{\circ}$ with the proper polarization configuration to generate one-dimension (1D) SOC.
Then we extend this case into a chiral 2D optical square Raman lattice to generate the SOC.

In Section 2, we introduce the chirality of spin motion in a static and AC magnetic field to better understand the physics of the chiral Raman coupling for SOC generated by chiral light-matter interaction.
We then present a scheme of generating chiral 1D Raman coupling for SOC by employing two Raman lasers at an intersecting angle 90$^{\circ}$ with the proper polarization configuration in Section 3 and extend this case into a chiral 2D optical square Raman lattice in Section 4.

\section{II. Chirality of spin motion in a static and AC magnetic field}

\begin{figure*}[!htb]
\includegraphics[width=6.6in]{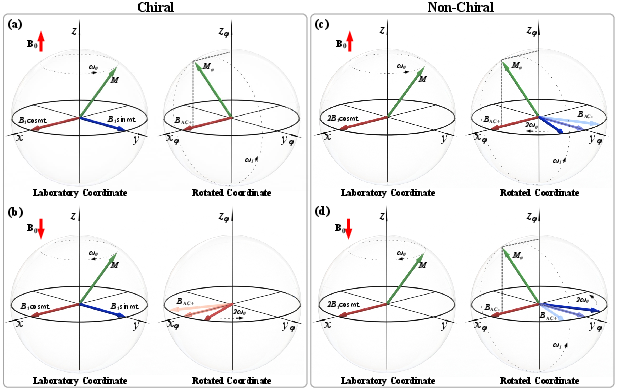}
\caption{(Color online). \textbf{Spin motion in a static and AC magnetic field.}
(a) and (b) Spin motion for the circular polarization of AC magnetic field $\textbf{\emph{B}}_{AC+}=$$B_{1}\cos \omega t \mathbf{\hat{e}_{x}}$+$ B_{1}\sin \omega t \mathbf{\hat{e}_{y}}$ with the strong, static magnetic field $B_{0}$ along the $+z$ (a) and $-z$ (b) respectively. When $B_{0}$ is along the $+z$ direction, the $\textbf{\emph{B}}_{AC+}$ in the rotating coordinate leads to the magnetization $\textbf{\emph{M}}_{\varphi}$ to rotate around $x_{\varphi}$ at the frequency of $\omega_{1}$ for $\omega=\omega_{0}$. When $B_{0}$ is along the $-z$ direction, the AC magnetic field $\textbf{\emph{B}}_{AC+}$ will be the high-speed rotation at the frequency of 2$\omega_{0}$ and its influence on the magnetization is negligible. This case is chiral since the chiral AC field induces the chiral interaction with spins. (c) and (d) Spin motion for the linear polarization of AC field $\textbf{\emph{B}}_{ACx}$=2$B_{1}\cos \omega t \mathbf{\hat{e}_{x}}$ with the strong, static magnetic field $B_{0}$ along the $+z$ (c) and $-z$ (d) respectively. When $B_{0}$ is along the $+z$ direction, only the component $\textbf{\emph{B}}_{AC+}$ in the rotating coordinate leads to the magnetization $\textbf{\emph{M}}_{\varphi}$ to rotate around $x_{\varphi}$ at the frequency of $\omega_{1}$ for $\omega=\omega_{0}$ and the component $\textbf{\emph{B}}_{AC-}$ is neglected due to the high-speed rotation at the frequency of 2$\omega_{0}$. When $B_{0}$ is reversed to the $-z$ direction, the role of two orthogonal circular components is reversed: the component $\textbf{\emph{B}}_{AC-}$ leads to the magnetization $\textbf{\emph{M}}_{\varphi}$ to rotate around $x_{\varphi}$ and the component $\textbf{\emph{B}}_{AC+}$ is neglected. Therefore, this case is non-chiral.
\label{Bloch}}
\end{figure*}

A spin-1/2 has a magnetic dipole moment $\mu_{s}$. When subjected to an external magnetic field $\textbf{\emph{B}}$, a dipole moment experiences a torque $\textbf{\emph{L}}$ according to
$\textbf{\emph{L}}=\mu_{s} \times \textbf{\emph{B}}$ (known as Larmor precession). Magnetization is defined as the vector sum of the spin magnetic moments $\mu_{s}$ per unit volume, denoted by the symbol \textbf{\emph{M}}
\begin{eqnarray}\label{eq:HH1}
\textbf{\emph{M}}=\frac{1}{V} \sum\limits_{k=1}^{N} \mu_{sk}
\end{eqnarray}
The spin with angular momentum \textbf{S} and magnetic moment $\mu_{s}$ is subjected to torque \textbf{L} under the external magnetic field \textbf{B}, which satisfies the equation
\begin{eqnarray} \label{eq:HH2}
\frac{d\mathbf{S}}{d\emph{t}}=\textbf{\emph{L}}=\mu_{s} \times \textbf{\emph{B}}
\end{eqnarray}
According to the formula $\mu_{s}=-\gamma \textbf{\emph{S}}$, Eq.~(\ref{eq:HH2}) can be written as
\begin{eqnarray}\label{eq:HH3}
\frac{d\mu_{s}}{d\emph{t}}=-\gamma \mu_{s} \times \textbf{\emph{B}}
\end{eqnarray}
where $\gamma$ is a proportionality constant called the gyromagnetic ratio. From the relation between \textbf{\emph{M}} and $\mu_{s}$, the magnetization vector expression of Bloch equation can be obtained as
\begin{eqnarray}\label{eq:HH4}
\frac{d\textbf{\emph{M}} }{d\emph{t}}=-\gamma \textbf{\emph{M}}  \times \textbf{\emph{B}}
\end{eqnarray}

Considering that a strong, static magnetic field $B_{0}$ along the $z$ direction is used to align spins.
When the spins initially are prepared along  the $z$ direction, an AC magnetic field $\textbf{\emph{B}}_{AC}$ (Radio-frequency field) is applied, which is perpendicular to the $z$ direction and then will rotate the spins away from $B_{0}$ called excitation.
Note that the influence on the magnetization is negligible when AC magnetic field $\textbf{\emph{B}}_{AC}$ is along the direction of $B_{0}$. Here, AC magnetic field $\textbf{\emph{B}}_{AC}$ has the polarization helicity $\sigma\in(-1,1)$.
The polarization helicity (the degree of circular polarization) is 0, $\pm1$ for linear, right- and left-hand circular polarization.
Therefore, the chiral AC magnetic field can induce the chiral interaction with atoms.

The chirality of spin motion in a static and AC magnetic field can be understood through the Bloch sphere in a laboratory and rotating coordinate systems as shown in Fig.~\ref{Bloch}.
At the strong, static magnetic field $B_{0}$ along the $+z$ direction of the laboratory coordinate, the spins precess in the direction of the external magnetic field at the Larmor frequency $\omega_{0}$=$\gamma B_{0}$.
So the magnetization remains stationary when we consider transforming the laboratory coordinate system into a rotating coordinate system ($x_{\varphi}$, $y_{\varphi}$, $z_{\varphi}$) that rotates around the $z$-axis at a frequency of $\omega_{0}$ as shown in the right panels of Fig.~\ref{Bloch}.
First, considering the circular polarization of AC magnetic field $\textbf{\emph{B}}_{AC+}=$$ B_{1}\cos \omega t ~\mathbf{\hat{e}_{x}}$+$ B_{1}\sin \omega t ~\mathbf{\hat{e}_{y}}$ with angular velocity $\omega$ as shown in Fig.~\ref{Bloch}(a), that is $\textbf{\emph{B}}$=$B_{1}\cos \omega t ~\mathbf{\hat{e}_{x}}$+$ B_{1}\sin \omega t ~\mathbf{\hat{e}_{y}}$+$ B_{0} \mathbf{\hat{e}_{z}}$.
So the Bloch equation can be written in laboratory coordinate
\begin{eqnarray}\label{eq:HH5}
        \frac{dM_{x}}{d\emph{t}}&=&\gamma (B_{1}M_{z}\sin \omega t-B_{0} M_{y}), \nonumber \\
        \frac{dM_{y}}{d\emph{t}}&=&\gamma (B_{0} M_{x}-B_{1}M_{z}\cos \omega t),  \\
 \frac{dM_{z}}{d\emph{t}}&=&\gamma (B_{1}M_{y}\cos \omega t-B_{1}M_{x}\sin \omega t). \nonumber
\end{eqnarray}
Bloch equation may be transformed into the rotating coordinate with the angular frequency $\omega$  with the transformation matrix shown as

\emph{\begin{eqnarray}\label{eq:HH6}
  \left(
      \begin{array}{ccc}
        \textbf{\emph{M}}_{\varphi x} \\
        \textbf{\emph{M}}_{\varphi y} \\
        \textbf{\emph{M}}_{\varphi z} \\
      \end{array}
    \right)=\left(
      \begin{array}{ccc}
        \cos \omega t & \sin \omega t & 0 \\
        -\sin \omega t & \cos \omega t & 0 \\
        0 & 0 & 1 \\
      \end{array}
    \right) \left(
      \begin{array}{ccc}
        \textbf{\emph{M}}_{x} \\
        \textbf{\emph{M}}_{y} \\
        \textbf{\emph{M}}_{z} \\
      \end{array}
    \right)
\end{eqnarray}}
Therefore, Bloch equation in the rotating coordinate is given as
\begin{eqnarray}\label{eq:HH7}
\frac{dM_{\varphi x}}{d\emph{t}}&=&-(\omega_{0}-\omega)M_{\varphi y},  \nonumber \\
\frac{dM_{\varphi y}}{d\emph{t}}&=&(\omega_{0}-\omega)M_{\varphi x}-\omega_{1}M_{\varphi z},   \\
\frac{dM_{\varphi z}}{d\emph{t}}&=&\omega_{1}M_{\varphi y}.  \nonumber
\end{eqnarray}
Here, $\omega_{1}$=$\gamma B_{1}$. When choose $\omega=\omega_{0}$, the AC magnetic field $\textbf{\emph{B}}_{AC+}$ always keeps along the $x_{\varphi}$ direction in the rotating coordinate, then the magnetization $\textbf{\emph{M}}_{\varphi}$ will rotate around $x_{\varphi}$ at the frequency of $\omega_{1}$ as shown in the right panel of Fig.~\ref{Bloch}(a).
The frequency $\omega_{1}$ that $\textbf{\emph{M}}_{\varphi}$ rotates around $B_{x_{\varphi}}$ is called the Rabi frequency.
If $B_{0}$ is reversed to the $-z$ direction, the AC magnetic field $B_{x_{\varphi}}$ will be the high-speed rotation at the frequency of 2$\omega_{0}$ as shown in Fig.~\ref{Bloch}(b).
Thus, the average time performance is close to zero, and the influence on the magnetization is negligible. This case is chiral since the chiral AC magnetic field induces the chiral interaction with spins.

Then considering the linear polarization of the AC magnetic field field $\textbf{\emph{B}}_{ACx}$=2$B_{1}\cos \omega t \mathbf{\hat{e}_{x}}$ applied in the $x$ direction of the laboratory coordinate system as shown in Fig.~\ref{Bloch}(c).
It can be decomposed into two orthogonal components $\textbf{\emph{B}}_{AC+}$=$B_{1}\cos \omega t \mathbf{\hat{e}_{x}}$+$B_{1}\sin \omega t \mathbf{\hat{e}_{y}}$ and $\textbf{\emph{B}}_{AC-}$=$B_{1}\cos \omega t \mathbf{\hat{e}_{x}}-B_{1}\sin \omega t \mathbf{\hat{e}_{y}}$ with the same amplitude, frequency and opposite directions of rotation.
When the strong, static magnetic field $B_{0}$ is along the $+z$ direction, only the component $\textbf{\emph{B}}_{AC+}$ in the rotating coordinate leads to the magnetization $\textbf{\emph{M}}_{\varphi}$ to rotate around $x_{\varphi}$ at the frequency of $\omega_{1}$ for $\omega=\omega_{0}$ and the component $\textbf{\emph{B}}_{AC-}$ is neglected due to the high-speed rotation at the frequency of 2$\omega_{0}$.
When $B_{0}$ is reversed to the $-z$ direction, the role of two orthogonal circular components is reversed: the component $\textbf{\emph{B}}_{AC-}$ leads to the magnetization $\textbf{\emph{M}}_{\varphi}$ to rotate around $x_{\varphi}$ and the component $\textbf{\emph{B}}_{AC+}$ is neglected as shown in Fig.~\ref{Bloch}(d). This case corresponds to non-chiral.

\section{III. Chiral Raman coupling for spin-orbit coupling}

As we all know, two Raman lasers can generate equivalently AC magnetic field $\textbf{\emph{B}}_{AC}$~\cite{Jessen1998PRA,Spielman2012NJP,Spielman2014RPP}. The effective magnetic field induced by the two Raman lasers $\textbf{\emph{E}}=\textbf{\emph{E}}_{1}e^{-i\omega_{L} t}+\textbf{\emph{E}}_{2}e^{-i(\omega_{L}+\delta\omega) t}$ can be wrote as
\begin{eqnarray}\label{eq:HH8}
\textbf{\emph{B}}^{eff}=\frac{i\mu_{v}}{\mu_{B} g_{J}}\textbf{\emph{E}}^{\ast}\times\textbf{\emph{E}}=\textbf{\emph{B}}_{DC}^{eff}+\textbf{\emph{B}}_{AC}^{eff},
\end{eqnarray}
where
\begin{eqnarray}\label{eq:HH9}
\textbf{\emph{B}}_{DC}^{eff}&=\frac{i\mu_{v}}{\mu_{B} g_{J}} (\textbf{\emph{E}}_{1}^{\ast}\times \textbf{\emph{E}}_{1}+\textbf{\emph{E}}_{2}^{\ast}\times \textbf{\emph{E}}_{2}),
\end{eqnarray}
\begin{eqnarray}\label{eq:HH10}
\textbf{\emph{B}}_{AC}^{eff}&=\frac{i\mu_{v}}{\mu_{B} g_{J}} (\textbf{\emph{E}}_{1}^{\ast}\times \textbf{\emph{E}}_{2}e^{-i \delta\omega t}+\textbf{\emph{E}}_{2}^{\ast}\times \textbf{\emph{E}}_{1}e^{i \delta\omega t}).
\end{eqnarray}
Here, $\omega_{L}$ is the frequency of the Raman lasers with the frequency difference $\delta\omega$, $\mu_{v}$ is the vector polarizability, $\mu_{B}$ is the Bohr magneton, $g_{J}$ is the electronic spin Land$\acute{e}$ g-factor.
From Eq.~(\ref{eq:HH9}), we can see that the first term $\textbf{\emph{B}}_{DC}^{eff}$ depends on the ellipticity of light, which corresponds to the effective DC magnetic field and is equivalent to the vector shift to generate linear Zeeman splitting (light shift proportional to $m_{F}$).
The $\textbf{\emph{B}}_{DC}^{eff}$ will add to the static bias field $\textbf{\emph{B}}_{DC}$ to act on atoms.
Therefore the influence on the magnetization is negligible when $\textbf{\emph{B}}_{DC}^{eff}$ is perpendicular to the direction of $B_{0}$.
The second term $\textbf{\emph{B}}_{AC}^{eff}$ corresponds to the effective AC magnetic field, in which only the components perpendicular to the direction of $B_{0}$ can drive Raman transitions between different energy levels $\delta m_{F}$=$\pm 1$. Here, $\delta m_{F}$=$\pm 1$ corresponds to the chiral transition. In this work, we only consider the second term $\textbf{\emph{B}}_{AC}^{eff}$ and neglect the $\textbf{\emph{B}}_{DC}^{eff}$.

The Hamiltonian including light-atom interaction as the effective magnetic field interaction is given as
\begin{eqnarray}\label{HH16}
\begin{aligned}
H=&\big[\frac{ \hat{\textbf{\emph{p}}}^{2}}{2m}+V(\textbf{r})\big]\hat{I}+\frac{\hbar \omega_{hf}}{2}\hat{\sigma}_{z}\\
&+\frac{\mu_{B}g^{m_{F}}_{m_{F'}}}{2}\textbf{\emph{B}}_{0}\cdot \hat{\mathbf{\sigma}}+\frac{\mu_{B}\eta^{m_{F}}_{m_{F'}}}{2}\textbf{\emph{B}}_{AC}^{eff}\cdot \hat{\mathbf{\sigma}},
\end{aligned}
\end{eqnarray}
where $\hat{I}$ is the identity operator and $\hat{\mathbf{\sigma}}$ is denoted with the three spin operators ($\hat{\sigma}_{z},\hat{\sigma}_{x},\hat{\sigma}_{y}$) or ($\hat{\sigma}_{z},\hat{\sigma}_{+},\hat{\sigma}_{-}$), $V(\textbf{r})$ is trapping potential, and $\hat{\sigma}_{\pm}=(\hat{\sigma}_{x}\pm i\hat{\sigma}_{y})/2$, $\eta^{m_{F}}_{m_{F'}}$ is a coupling constant.
For some special geometry structure of the  Raman lasers, the effective AC magnetic field $\textbf{\emph{B}}_{AC}^{eff}$ is spatial dependent. As a result, one can realize spin-orbit coupling from the last term of Eq.~(\ref{HH16}). \\

\begin{figure}[t]
\includegraphics[width=3.2in]{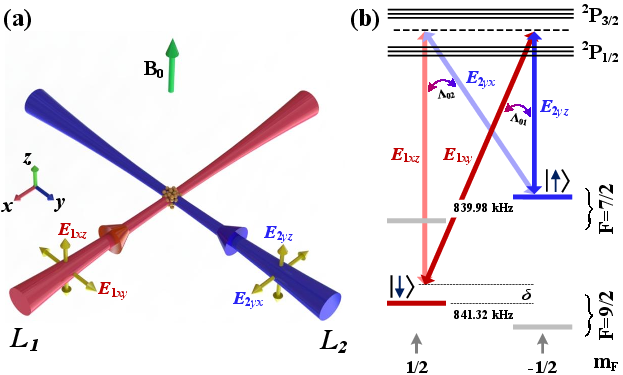}
\caption{(Color online). \textbf{Schematic of experimental scheme of 1D Raman coupling for SOC for $\delta m_{F}=\pm 1$ and the energy level of $^{40}$K.}
(a)~The Raman lasers $L_{1}$ (red) and $L_{2}$ (blue) propagate along the $x$ and $y$ axis respectively.
Here, the Raman coupling is induced by ($E_{1xz}$,$E_{2yx}$) and ($E_{2yz}$,$E_{1xy}$).
The external magnetic field $\textbf{\emph{B}}_{0}$ along the $z$ direction generates the Zeeman splitting and the quantization axis of the system.
(b)~Two hyperfine spin states are coupled with the electronic excited states through two Raman lasers $L_{1}$  and $L_{2}$. Here, the Raman lasers drive the chiral transition $\delta m_{F}=\pm 1$ between $|F=7/2, m_{F}=-1/2\rangle$$\equiv$$\left\vert \uparrow \right\rangle$ and $|F=9/2, m_{F}=1/2\rangle$$\equiv$$\left\vert \downarrow \right\rangle$.
\label{setup}}
\end{figure}

\subsection{A. Chiral 1D Raman coupling for spin-orbit coupling}

Here, we employ two Raman lasers at an intersecting angle 90$^{\circ}$ in the x-y plane to generate chiral 1D Raman coupling for SOC. A homogeneous magnetic biased field $B_{0}$=2.7 G is applied in the $z$ axis (gravity direction shown in Fig.~\ref{setup}), which defines the magnetic quantization axis and gives a Zeeman splitting. Two hyperfine Zeeman states $|F=7/2, m_{F}=-1/2\rangle$$\equiv$$\left\vert \uparrow \right\rangle$ and $|F=9/2, m_{F}=1/2\rangle$$\equiv$$\left\vert \downarrow \right\rangle$ are chosen as the two spin states for Raman transition. The bichromatic light field of the Raman scheme is written as
\begin{eqnarray}\label{HH11}
\begin{aligned}
\mathbf{E}=\mathbf{E_{1x}}+\mathbf{E_{2y}}
\end{aligned}
\end{eqnarray}
where
\begin{eqnarray}\label{HH12}
\begin{aligned}
\mathbf{E_{1x}}&=(\mathbf{\hat{e}_{z}}E_{1xz}+\mathbf{\hat{e}_{y}}
E_{1xy}e^{i\varphi_{1}})e^{-ik_{R}x}e^{-i(\omega_{L}+\delta\omega)t}& \\
\mathbf{E_{2y}}&=(\mathbf{\hat{e}_{z}}E_{2yz}+\mathbf{\hat{e}_{x}}
E_{2yx}e^{i\varphi_{2}})e^{-ik_{R}y}e^{-i\omega_{L} t}&
\end{aligned}
\end{eqnarray}
Here the $\varphi_{1}$ and $\varphi_{2}$ are the initial phase of two orthogonally polarized components of the laser beam 1 and 2 respectively, which determine the polarization helicity of the input Raman laser beam 1 and 2, as shown in Fig.~\ref{setup}(a). $k_{R}=2\pi/\lambda$ is the wave number  of the Raman laser with $\lambda$ being the wavelength. The frequency difference $\delta\omega=\omega_{hf}+\mu_{B}g^{m_{F}}_{m_{F'}}B_{0}/\hbar+\delta$ differes by a small detuning $\delta$ from the hyperfine energy splitting $\omega_{hf}$ and the linear Zeeman shift between $\delta m_{F}=\pm 1$. Here, $g^{m_{F}}_{m_{F'}}$ depends on the hyperfine Land\'{e} g-factors $g_{F}$ and $g_{F'}$.

Then, the effective AC magnetic field can be expressed as ~\cite{Spielman2014RPP}

\begin{eqnarray}\label{HH13}
\begin{aligned}
\textbf{\emph{B}}_{AC}^{eff}=\textbf{\emph{B}}_{AC-el}^{eff}e^{i\delta\omega t}e^{ik_{R}(x-y)}+\textbf{\emph{B}}_{AC-el}^{eff \dagger}e^{-i\delta\omega t}e^{-ik_{R}(x-y)},
\end{aligned}
\end{eqnarray}
where

\begin{eqnarray}\label{HH14}
\begin{aligned}
&\textbf{\emph{B}}_{AC-el}^{eff}=i\Lambda_{01}e^{-i\varphi_{1}}\mathbf{\hat{e}_{x}}
+i\Lambda_{02}e^{i\varphi_{2}}\mathbf{\hat{e}_{y}}-i\Lambda_{03}e^{i(\varphi_{2}-\varphi_{1})}\mathbf{\hat{e}_{z}}.
\end{aligned}
\end{eqnarray}
Here,
\begin{eqnarray}\label{HH15}
\begin{aligned}
&\Lambda_{01}=\frac{u_{v}}{\mu_{B}g_{J}}E_{1xy}E_{2yz}, \\
&\Lambda_{02}=\frac{u_{v}}{\mu_{B}g_{J}}E_{1xz} E_{2yx}, \\
&\Lambda_{03}=\frac{u_{v}}{\mu_{B}g_{J}}E_{1xy}E_{2yx}.
\end{aligned}
\end{eqnarray}
Since the static bias field $\textbf{\emph{B}}_{0}$ is along the $z$ axis and Raman coupling drives transitions between different energy levels $\delta m_{F}=\pm 1$, the term $-i\Lambda_{03}e^{i(\varphi_{2}-\varphi_{1})}\mathbf{\hat{e}_{z}}$ of $\textbf{\emph{B}}_{AC}^{eff}$ can be neglected. Here, $\textbf{\emph{B}}_{AC-el}^{eff}$ and  $\textbf{\emph{B}}_{AC-el}^{eff\dagger}$ represent the ellipticity of the effective AC magnetic field, which is determined by the phases $\varphi_{1}$ and $\varphi_{2}$.

The time and space-dependence $e^{i[\delta\omega t+k_{R}(x-y)]}$ of $\textbf{\emph{B}}_{AC}^{eff}$ can be eliminated via the unitary transformation $S=e^{i[\delta\omega t+k_{R}(x-y)]\hat{\sigma}_{z}/2}$ and the rotating wave approximation. The resulting Hamiltonian can be represented as
\begin{eqnarray} \label{HH17}
\begin{aligned}
H_{R}=&\big[\frac{\hat{\textbf{\emph{p}}}^{2}}{2m}+V(r)\big]\hat{I}-\frac{\hbar k_{R}(p_{x}-p_{y})}{2m}\hat{\sigma}_{z} \\
&+\mathbf{\Omega}\cdot \hat{\mathbf{\sigma}}+\frac{(\hbar k_{R})^{2}}{4m}
\end{aligned}
\end{eqnarray}
where
\begin{eqnarray}\label{HH18}
\begin{aligned}
&\mathbf{\Omega}\cdot \hat{\mathbf{\sigma}}=\Omega_{z}\hat{\sigma}_{z}+\Omega\hat{\sigma}_{-}+\Omega^{\dagger}\hat{\sigma}_{+},& \\
&\Omega_{z}=-\frac{\hbar\delta}{2},& \\
&\Omega=\frac{\mu_{B}\eta^{m_{F}}_{m_{F'}}}{2}\textbf{\emph{B}}_{AC-el}^{eff}\cdot
(\mathbf{\hat{e}_{x}}+i\mathbf{\hat{e}_{y}}),& \\
&\Omega^{\dagger}=\frac{\mu_{B}\eta^{m_{F}}_{m_{F'}}}{2}\textbf{\emph{B}}_{AC-el}^{eff\dagger}\cdot
(\mathbf{\hat{e}_{x}}-i\mathbf{\hat{e}_{y}}).&
\end{aligned}
\end{eqnarray}
Then we can obtain the simplified Hamiltonian as
\begin{eqnarray}\label{HH19}
\begin{aligned}
H_{R}=&[\frac{\hat{\textbf{\emph{p}}}^{2}}{2m}+V(r)]\hat{I}-\frac{\hbar k_{R}(p_{x}-p_{y})}{2m}\hat{\sigma}_{z}-\frac{\hbar\delta}{2}\hat{\sigma}_{z} \\ &+\frac{\mu _{B}\eta_{m_{F^{\prime }}}^{m_{F}}}{2}\Lambda _{01}\left[ \sin
\varphi _{1}\sigma _{x}+\cos \varphi _{1}\sigma _{y}\right] \\ &-\frac{\mu
_{B}\eta_{m_{F^{\prime }}}^{m_{F}}}{2}\Lambda _{02}\left[ \cos \varphi
_{2}\sigma _{x}+\sin \varphi _{2}\sigma _{y}\right].
\end{aligned}
\end{eqnarray}

When $B_{0}$ is reversed to the $-z$ direction, the role of two orthogonal circular components of the effective AC magnetic field is reversed
\begin{eqnarray}\label{HH20}
\begin{aligned}
&\Omega=\frac{\mu_{B}\eta^{m_{F}}_{m_{F'}}}{2}\textbf{\emph{B}}_{AC-el}^{eff\dagger}\cdot
(\mathbf{\hat{e}_{x}}+i\mathbf{\hat{e}_{y}}),& \\
&\Omega^{\dagger}=\frac{\mu_{B}\eta^{m_{F}}_{m_{F'}}}{2}\textbf{\emph{B}}_{AC-el}^{eff}\cdot
(\mathbf{\hat{e}_{x}}-i\mathbf{\hat{e}_{y}}).&
\end{aligned}
\end{eqnarray}
The Hamiltonian in Eq.~(\ref{HH17}) describes a 1D SOC consisting of equal sum of Rashba and Dresselhaus terms. This system is non-trivial due to the non-commutativity between the Abelian gauge potential $\mathcal{A}=\hbar k_{R}(\mathbf{\hat{e}_{x}}-\mathbf{\hat{e}_{y}})/2$ and an additional Raman coupling term $\mathbf{\Omega}\cdot \hat{\mathbf{\sigma}}$, and leads to lots of interesting phenomena studied both theoretically~\cite{Spielman2009PRA,Ho2011PRL,Spielman2012PRL,Stringari2013PRL,Han2018PRL} and experimentally~\cite{Jung2012PRL,Engels2014nc,Pu2016PRA,Meng2016PRL,Ketterle2016PRL,Wolfgang2017Nature,Chuanwei2019nc}.

\begin{figure*}[!htb]
\includegraphics[width=5.5in]{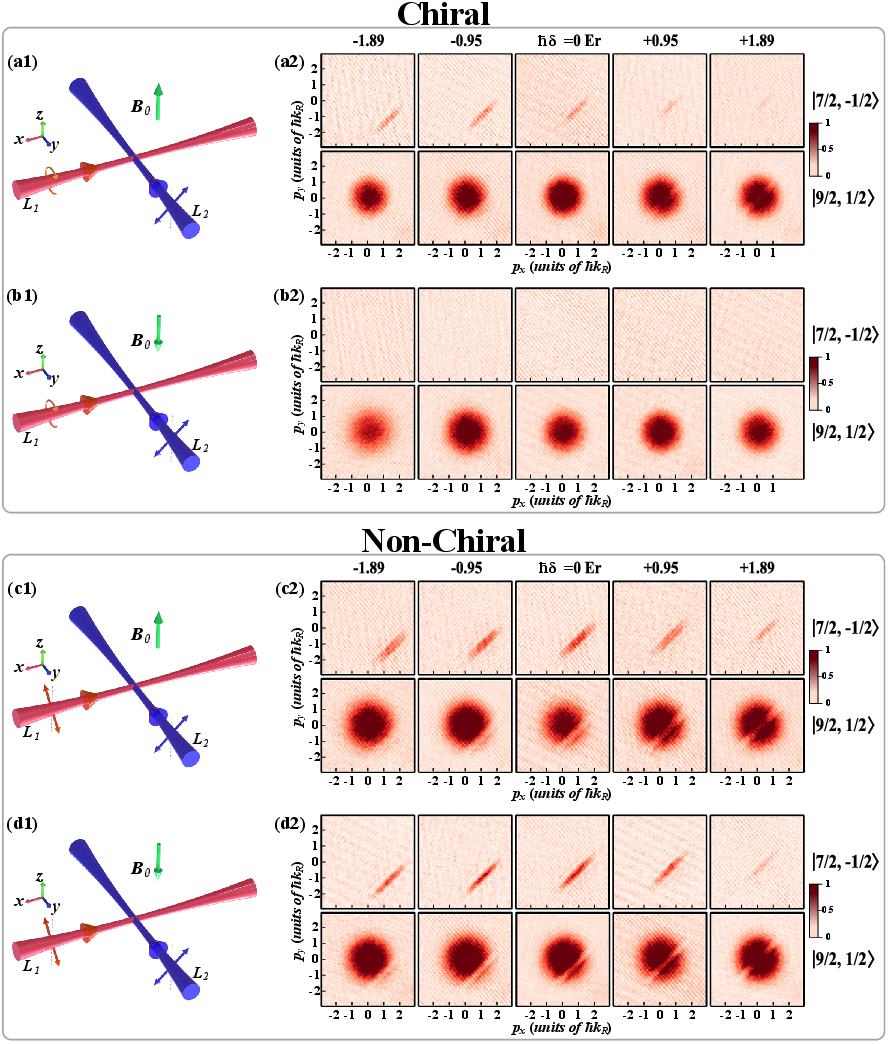}
\caption{(Color online). \textbf{Raman spectroscopy of 1D Raman coupling for SOC for $\delta m_{F}$=$\pm 1$.}
(a) and (b) Chiral 1D Raman coupling for SOC. The external magnetic field $\textbf{\emph{B}}_{0}$ is along the $+z$ direction for (a) and $-z$ direction for (b), respectively. (a1) and (b1)~The polarization of the Raman laser $L_{1}$ along $x$ is circular while $L_{2}$ along $y$ is linear at a $\pi/4$ angle to the vertical direction. (a2) and (b2)~The absorption images of two hyperfine spin states after 12 ms TOF for the different Raman two-photon-detuning $\delta$.
(c) and (d) Nonchiral 1D Raman coupling for SOC. (c1) and (d1)~The polarization of the Raman laser $L_{1}$ and $L_{2}$ is linear and orthogonal at the $\pm\pi/4$ angle to the vertical direction. (c2) and (d2)~The absorption images of two hyperfine spin states after 12 ms TOF for the different Raman two-photon detuning $\delta$.
\label{Raman-red}}
\end{figure*}

To simplify without losing generality, we choose $|E_{1xy}|=|E_{1xz}|=|E_{2yz}|=|E_{2yx}|$ ($\Lambda_{01}=\Lambda_{02}=\Omega_{0}/\mu_{B}\eta^{m_{F}}_{m_{F'}}$).
For the setting with $(\varphi_{1},\varphi_{2})=(-\pi/2, 0)$ hown in Fig.~\ref{Raman-red}(a1), $\Omega=\Omega^{\dagger}=\Omega_{0}$ and $\mathbf{\Omega}\cdot \hat{\mathbf{\sigma}}= -\hat{\sigma}_{z}\delta/2-\Omega_{0}\hat{\sigma}_{x}$.
This corresponds to generate the effective circular polarization (see Eq.~(\ref{HH14})) of the AC magnetic field and leads to the Raman coupling term $-\Omega_{0}\hat{\sigma}_{x}$.
 When $B_{0}$ is reversed to the $-z$ direction (Fig.~\ref{Raman-red}(b1)), this is similar to the setting with $(\varphi_{1},\varphi_{2})=(\pi/2, 0)$ to generate the orthogonal effective circular polarization of the AC magnetic field. Then we have $\mathbf{\Omega}\cdot \hat{\mathbf{\sigma}}= -\hat{\sigma}_{z}\delta/2$, which means that the Raman coupling disappears.
Therefore this case corresponds to chiral 1D Raman coupling for SOC since the effective chiral RF field induces the chiral interaction with atoms.

For the setting with $(\varphi_{1},\varphi_{2})=(0,0)$ shown in Fig.~\ref{Raman-red}(c1), $\Omega=-\Omega_{0}(1- i)/2$, $\Omega^{\dagger}=-\Omega_{0}(1+ i)/2$ and $\mathbf{\Omega}\cdot \hat{\mathbf{\sigma}}= -\hat{\sigma}_{z}\delta/2-\Omega_{0}(\hat{\sigma}_{x}-\hat{\sigma}_{y})/2$.
This corresponds to generate the effective linear polarization of the RF field and leads to the Raman coupling term $-\Omega_{0}(\hat{\sigma}_{x}-\hat{\sigma}_{y})/2$.
When $B_{0}$ is reversed to the $-z$ direction shown in Fig.~\ref{Raman-red}(d1), $\Omega=\Omega_{0}(1- i)/2$, $\Omega^{\dagger}=\Omega_{0}(1+ i)/2$ and $\mathbf{\Omega}\cdot \hat{\mathbf{\sigma}}= -\hat{\sigma}_{z}\delta/2+\Omega_{0}(\hat{\sigma}_{x}-\hat{\sigma}_{y})/2$.
Therefore this case corresponds to non-chiral 1D Raman coupling for SOC since the similar SOC exists in $\pm z$ directions of the magnetic quantization axis.

Furthermore, we consider the special case of driving Raman transition between different energy levels with $\delta m_{F}=0$, which corresponds to the nonchiral transition, such as two hyperfine Zeeman states $|F=7/2, m_{F}=1/2\rangle$$\equiv$$\left\vert \uparrow \right\rangle$ and $|F=9/2, m_{F}=1/2\rangle$$\equiv$$\left\vert \downarrow \right\rangle$.
Only the component $\Lambda_{03}e^{-i(\varphi_{2}-\varphi_{1})}\mathbf{\hat{e}_{z}}$ of the effective AC magnetic field $\textbf{\emph{B}}_{AC}^{eff}$ (Eq.~(\ref{HH14})) parallel to the static bias field $\textbf{\emph{B}}_{0}$ can drive transition $\delta m_{F}=0$ and the components of $\Lambda_{01}e^{-i\varphi_{1}}\mathbf{\hat{e}_{x}}$ and $\Lambda_{02}e^{i\varphi_{2}}\mathbf{\hat{e}_{y}}$ perpendicular to the static bias field $\textbf{\emph{B}}_{0}$ are neglected. Therefore, $\Omega$ and $\Omega^{\dagger}$ in Eq.~(\ref{HH18}) are expressed as
\begin{eqnarray}\label{HH21}
\begin{aligned}
&\Omega=\frac{\mu_{B}\eta^{m_{F}}_{m_{F'}}}{2}\textbf{\emph{B}}_{AC-el}^{eff}\cdot
\mathbf{\hat{e}_{z}},& \\
&\Omega^{\dagger}=\frac{\mu_{B}\eta^{m_{F}}_{m_{F'}}}{2}\textbf{\emph{B}}_{AC-el}^{eff\dagger}\cdot
\mathbf{\hat{e}_{z}}.&
\end{aligned}
\end{eqnarray}
Here, the amplitudes of $\Omega$ and $\Omega^{\dagger}$ are independent of the relative phase $(\varphi_{1},\varphi_{2})$.
When $B_{0}$ is reversed to the $-z$ direction, $\Omega$ and $\Omega^{\dagger}$ remain unchanged according to Eq.~(\ref{HH21}). Therefore this case corresponds to non-chiral 1D Raman coupling for SOC since the similar SOC exists in $\pm z$ directions of the magnetic quantization axis.
It should be noted that the non-chiral Raman coupling with $\delta m_{F}=0$ can be experimentally observed both in 1D and 2D. For simplicity, we only present the observation for 2D in the following, which is sufficient and more representative for verifying the theoretical analysis.

The  experimental setup of $^{87}$Rb and $^{40}$K Bose-Fermi mixture is presented in detail with our previous works~\cite{Chai2012Acta,Huang2022CPB,Ding2024CPB}.
The experiment starts with the preparation of a degenerate Fermi gas of $^{40}$K atoms in the state $|F=9/2,m_{F}=9/2\rangle$ in a crossed optical dipole trap.
Around $N=3\times10^{6}$ ultracold $^{40}$K atoms are prepared at a temperature of $0.3T_F$ using sympathetic cooling by $^{87}$Rb, where the Fermi temperature is defined by $T_F=\hbar \bar{\omega} (6N)^{1/3}/k_B$.
Here $\bar{\omega} =(\omega_x\omega_y\omega_z)^{1/3}$ $\simeq2$$\pi\times80$ Hz is the geometric mean of the optical trap in our experiment, $N$ is the particle number of $^{40}$K atoms, and $k_B$ is the Boltzmann's constant. The remaining $^{87}$Rb atoms are optically removed by applying a resonant laser beam pulse (780 nm) for 0.03 ms without heating and losing $^{40}$K atoms.
The initial spin state $|F=9/2, m_{F}=1/2\rangle$$\equiv$$\left\vert \downarrow \right\rangle$ is prepared in a rapid adiabatic passage induced by sweeping a Radio frequency (RF) field from $|F=9/2, m_{F}=9/2\rangle$ in 50 ms.

A pair of Raman lasers is located at the tune-out wavelength $\lambda$=768.98 nm~\cite{Alexander2017PRA}, which is extracted from a CW Ti-sapphire single-frequency laser. Therefore, the Raman laser approaches zero AC-stark energy shift for free atoms~\cite{Bian2022PRA}. The recoil energy $E_{R}=(\hbar k_{R})^{2}/2m$=$h\times8.44$ kHz is taken as the natural energy units. The Raman laser beam $L_{1}$ is frequency-shifted around +380.808 MHz by a single-pass acousto-optic modulator (AOM), and $L_{2}$ is frequency-shifted -226.25$\times$4 MHz through two AOMs in double-pass configuration respectively, which match the hyperfine transition frequency $\omega_{hf}=1.28$ GHz.

Here, we employ the momentum-resolved Raman spectroscopy of an ultracold Fermi gas to check the chiral Raman coupling for SOC.
We apply a Raman laser
pulse with the duration time of 35 $\mu$s and measure the spin population for
different frequency differences $\delta\omega$ of the Raman lasers.
After the Raman pulse, we immediately turn off
the optical trap and the homogeneous magnetic field, let the atoms free expand in 12 ms and take the time-of-flight
(TOF) absorption image.
The absorption images of two different hyperfine states for
different frequency differences $\delta\omega$ of the Raman lasers are shown in Fig.~\ref{Raman-red}.
We can see that only atoms in certain momentum state are transferred from $|F=9/2, m_{F}=1/2\rangle$$\equiv$$\left\vert \downarrow \right\rangle$ to $|F=7/2, m_{F}=-1/2\rangle$$\equiv$$\left\vert \uparrow \right\rangle$, which is determined by the frequency difference of the Raman lasers shown in Fig.~\ref{Raman-red}.
It shows the momentum-resolved features of Raman spectroscopy.

Consider that the polarization of the Raman laser $L_{1}$ along $x$ is chosen as circular while $L_{2}$ along $y$ as linear at a $\pi/4$ angle to the vertical direction as shown in Fig.~\ref{Raman-red}(a1).
This case corresponds to $|E_{1xy}|=|E_{1xz}|=|E_{2yz}|=|E_{2yx}|$ and the setting with $(\varphi_{1},\varphi_{2})=(-\pi/2, 0)$ to generate the effective circular polarization of RF field.
The absorption images for the momentum-resolved Raman spectroscopy in Fig.~\ref{Raman-red}(a2) demonstrate the existence of the Raman coupling $\Omega_{0}\hat{\sigma}_{x}$.
In contrast, when $B_{0}$ is reversed to the $-z$ direction as shown in Fig.~\ref{Raman-red}(b1) (or the polarization of $L_{1}$ is changed into the orthogonal circular state) which corresponds to the setting with $(\varphi_{1},\varphi_{2})=(\pi/2, 0)$ to generate the orthogonal effective circular polarization of AC magnetic field, there is no Raman coupling as shown in Fig.~\ref{Raman-red}(b2).
Therefore the chiral 1D Raman coupling for SOC is realized experimentally by the chiral interaction between the effective AC magnetic field and atoms.

\begin{figure*}[!htb]
\includegraphics[width=5.5in]{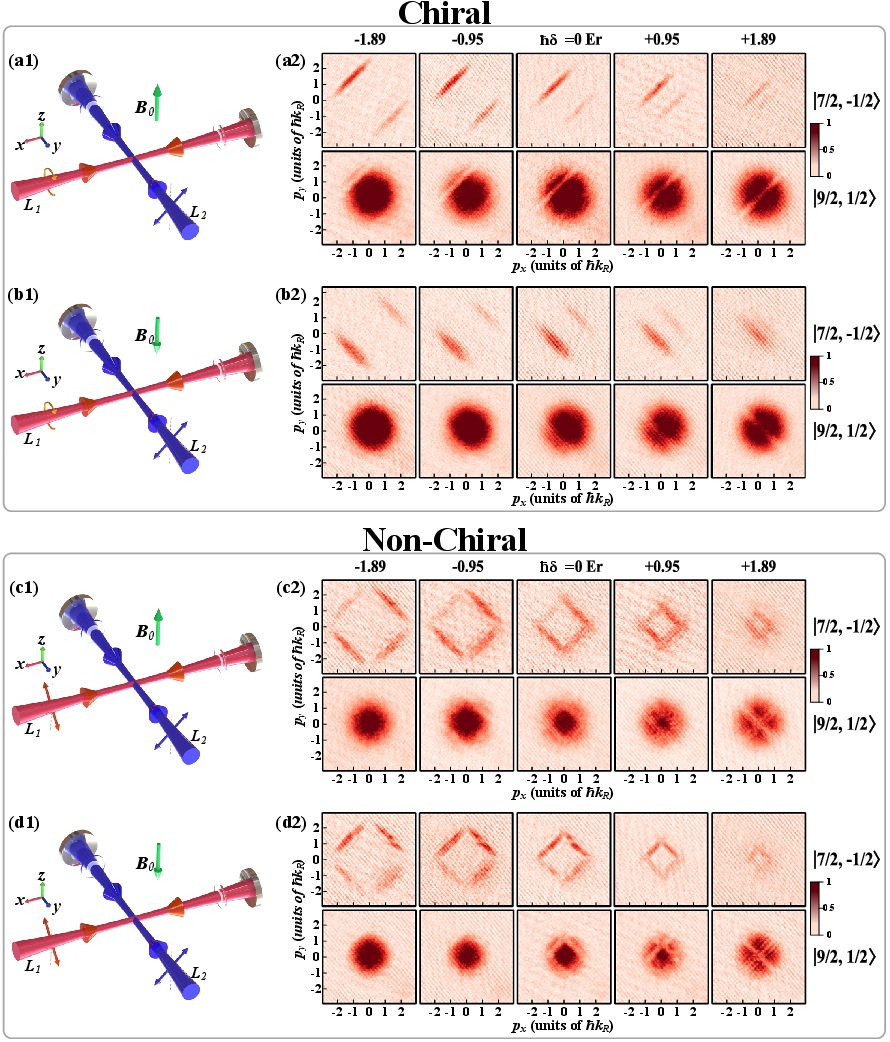}
\caption{(Color online). \textbf{Raman spectroscopy of 2D Raman coupling for SOC in an optical square Raman lattice for $\delta m_{F}=\pm 1$.}
(a) and (b) 2D chiral Raman coupling for SOC. The external magnetic field $\textbf{\emph{B}}_{0}$ is along the $+z$ direction for (a) and $-z$ direction for (b), respectively. (a1) and (b1)~The polarization of the Raman laser $L_{1}$ along $x$ is circular while $L_{2}$ along $y$ is linear at a $\pi/4$ angle to the vertical direction. (a2) and (b2)~The absorption images of two hyperfine spin states after 12 ms TOF for the different Raman two-photon detuning $\delta$.
(c) and (d) 2D nonchiral Raman coupling for SOC. (c1) and (d1)~The polarization of the Raman laser $L_{1}$ and $L_{2}$ is linear and orthogonal at the $\pm\pi/4$ angle to the vertical direction. (c2) and (d2)~The absorption images of two hyperfine spin states after 12 ms TOF for the different Raman two-photon detuning $\delta$.
\label{lattice-red}}
\end{figure*}

Consider that the polarization of the Raman laser $L_{1}$ and $L_{2}$ is linear and orthogonal at the $\pm\pi/4$ angle to the vertical direction respectively as shown in Fig.~\ref{Raman-red}(c1) and (d1).
This case corresponds to the setting with $(\varphi_{1},\varphi_{2})=(0,0)$ to generate the effective linear polarization of the AC magnetic field.
We can see that there always are Raman coupling in $\pm z$ directions of the magnetic quantization axis from the absorption images for the momentum-resolved Raman spectroscopy in Fig.~\ref{Raman-red}(c2) and (d2).
Therefore this case corresponds to non-chiral 1D Raman coupling for SOC.

\subsection{B. Chiral 2D Raman coupling for SOC in an optical square Raman lattice}

The two Raman lasers $\mathbf{E_{1x}}$ and $\mathbf{E_{2y}}$ are incident from x and y directions respectively and retro-reflected by two mirrors. The bichromatic light field of the Raman lattice scheme is written as
\begin{eqnarray}\label{HH22}
\begin{aligned}
\mathbf{E}=\mathbf{E_{1x}}+\mathbf{E_{2y}},
\end{aligned}
\end{eqnarray}
where

\begin{eqnarray}\label{HH23}
\begin{aligned}
\mathbf{E_{1x}}  =& 2[ \mathbf{\hat{e}_{z}}E_{1xz}\cos\left(
k_{R}x\right)  \\
&+\mathbf{\hat{e}_{y}}E_{1xy}e^{i\varphi_{1}}e^{i\frac{\alpha}{2}}\cos\left(
k_{R}x+\frac{\alpha}{2}\right)]   e^{-i\left(  \omega_{L}+\delta\omega\right)  t},\\
\mathbf{E_{2y}}  =& 2[ \mathbf{\hat{e}_{z}} E_{2yz}\cos\left(
k_{0}y\right) \\
&+\mathbf{\hat{e}_{x}} E_{2yx}e^{i\varphi_{2}}e^{i\frac{\beta}{2}}\cos\left(
k_{R}y+\frac{\beta}{2}\right) ]  e^{-i\omega_{L} t}.
\end{aligned}
\end{eqnarray}
Here, $\alpha$ ($\beta$) is the extra relative phase difference between two orthogonal polarization components acquired by $E_{1xy}$ ($E_{2yx}$) via a wave plate before and after the retro-reflected mirror.

\begin{figure*}[!htb]
\includegraphics[width=6.8in]{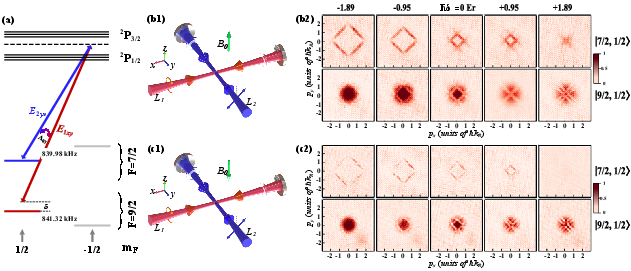}
\caption{(Color online). \textbf{Raman spectroscopy of 2D Raman coupling for SOC in an optical square Raman lattice for $\delta m_{F}$=$0$.}
(a)~Two Raman lasers drive the nonchiral Raman transition $\delta m_{F}$=$0$ between two hyperfine spin states $|F=7/2, m_{F}=1/2\rangle$ and $|F=9/2, m_{F}=1/2\rangle$. (b1) and (c1)~The polarization of $l_{L}$ along $x$ is circular while the Raman laser $L_{2}$ along $y$ is linear at a $\pi/4$ angle to the vertical direction.
The external magnetic field $\textbf{\emph{B}}_{0}$ is along the $+z$ (b1) and $-z$ direction (c1).
(b2) and (c2)~The absorption images of two hyperfine spin states after 12 ms TOF for different Raman frequency detuning $\delta$.
\label{different-transition}}
\end{figure*}

Then, the effective AC magnetic field can be expressed as
\begin{eqnarray}\label{HH24}
\begin{aligned}
\textbf{\emph{B}}^{\mathrm{eff}}_{AC}=\textbf{\emph{B}}^{\mathrm{eff}}_{AC-el}e^{i\delta\omega t}+\textbf{\emph{B}}^{eff\dagger}_{AC-el}e^{-i\delta\omega t}
\end{aligned}
\end{eqnarray}
where
\begin{eqnarray}\label{HH25}
\begin{aligned}
\textbf{\emph{B}}^{\mathrm{eff}}_{AC-el}  & =iM_{x}e^{-i\left(  \varphi_{1}+\alpha
/2\right)  }\mathbf{\hat{e}_{x}}+iM_{y}e^{i\left( \varphi_{2}+\beta/2\right)  }%
\mathbf{\hat{e}_{y}} \\
&-iM_{z}e^{i\left( \varphi_{2}-\varphi_{1}+\beta/2-\alpha
/2\right)  }%
\mathbf{\hat{e}_{z}}
\end{aligned}
\end{eqnarray}
Here,
\begin{eqnarray}\label{HH26}
\begin{aligned}
M_{x}  & =\Lambda_{01}\cos\left(  k_{R}x+\alpha/2\right)  \cos\left(
k_{0}y\right),  \\
M_{y}  & =\Lambda_{02}\cos\left(  k_{R}y+\beta/2\right)  \cos\left(
k_{0}x\right),\\
M_{z}  & =\Lambda_{03}\cos\left(  k_{R}x+\alpha/2\right) \cos\left(  k_{R}y+\beta/2\right)  .
\end{aligned}
\end{eqnarray}
and
\begin{eqnarray}\label{HH27}
\begin{aligned}
\Lambda_{01}  & =4\frac{u_{v}}{\mu_{B}g_{J}}E_{1xy}E_{2yz},\\
\Lambda_{02}  & =4\frac{u_{v}}{\mu_{B}g_{J}}E_{2yx}E_{1xz}, \\
\Lambda_{03}  & =4\frac{u_{v}}{\mu_{B}g_{J}}E_{1xy}E_{2yx}&.%
\end{aligned}
\end{eqnarray}

The time-dependence $e^{i\delta\omega t}$ of $\textbf{\emph{B}}_{AC}^{eff}$ can be eliminated via the unitary transformation $S=e^{i\delta\omega t\hat{\sigma}_{z}/2}$ and the rotating wave approximation. The resulting Hamiltonian can be represented as
\begin{eqnarray} \label{HH17}
\begin{aligned}
H_{R}=&\big[\frac{\hat{\textbf{\emph{p}}}^{2}}{2m}+V(r)\big]\hat{I} +\mathbf{\Omega}\cdot \hat{\mathbf{\sigma}}
\end{aligned}
\end{eqnarray}
where
\begin{eqnarray}\label{HH18}
\begin{aligned}
&\mathbf{\Omega}\cdot \hat{\mathbf{\sigma}}=\Omega_{z}\hat{\sigma}_{z}+\Omega\hat{\sigma}_{-}+\Omega^{\dagger}\hat{\sigma}_{+},& \\
&\Omega_{z}=-\frac{\hbar\delta}{2},& \\
&\Omega=\bar{M}_{x}\left[ \sin \left( \varphi _{1}+\alpha /2\right) +i\cos
\left( \varphi _{1}+\alpha /2\right) \right]& \\
&-\bar{M}_{y}\left[ \cos \left(\varphi _{2}+\beta /2\right) +i\sin \left( \varphi _{2}+\beta /2\right) %
\right],&
\end{aligned}
\end{eqnarray}
with $\bar{M}_{x/y}=\mu_{B}\eta^{m_{F}}_{m_{F'}}{M}_{x/y}/2$.

\begin{figure*}[!htb]
\includegraphics[width=6.8in]{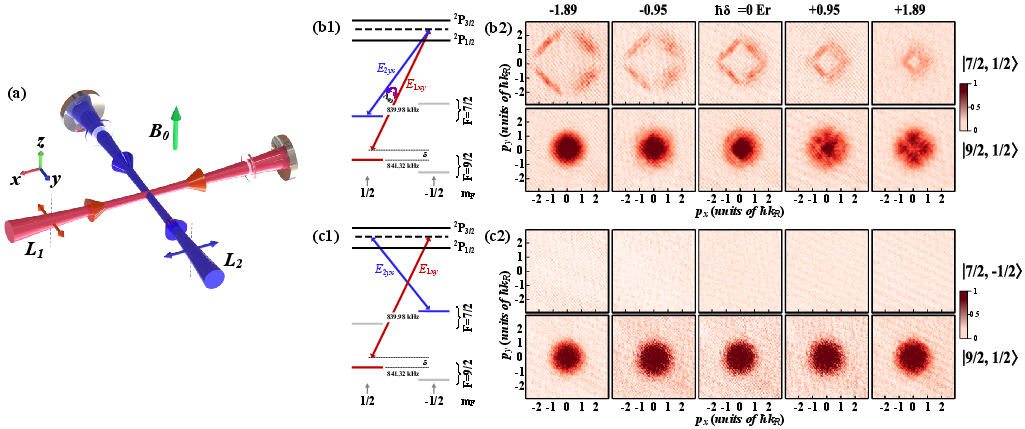}
\caption{(Color online). \textbf{Raman spectroscopy of 2D Raman coupling for SOC in an optical square Raman lattice comparing $\delta m_{F}$=$0$ and  $\pm1$ with the horizontally linear polarisation.}
(a)~The polarization of Raman laser $L_{1}$ along $x$ and $L_{2}$ along $y$ all have the horizontally linear polarisation to generate square lattice and Raman coupling respectively.
The external magnetic field $\textbf{\emph{B}}_{0}$ is along the $+z$.
(b1) and (c1)~Two Raman lasers drive the Raman transition $\delta m_{F}$=$0$ ($|9/2, 1/2\rangle$$\rightarrow$$|7/2, 1/2\rangle$) and $\delta m_{F}$=$\pm1$ ($|9/2, 1/2\rangle$$\rightarrow$$|7/2, -1/2\rangle$), respectively. (b2) and (c2)~The absorption images of two hyperfine spin states after 12 ms TOF for different Raman frequency detuning $\delta$.
\label{paipai}}
\end{figure*}

We consider that the $\lambda/4$ wave plate is placed in the front of both retro-reflected mirrors which corresponds to the extra relative phase difference between two orthogonal polarization $\alpha=\beta=\pi$.
For the setting with $(\varphi_{1},\varphi_{2})=(-\pi/2, 0)$ as shown in Fig.~\ref{lattice-red}(a1), in which the polarization of the Raman laser $L_{1}$ along $x$ is chosen as circular while $L_{2}$ along $y$ as linear at a $\pi/4$ angle to the vertical direction, $\Omega=i(\bar{M}_{x}-\bar{M}_{y})$ and $\mathbf{\Omega}\cdot \hat{\mathbf{\sigma}}=-\hat{\sigma}_{z}\delta/2+(\bar{M}_{x}-\bar{M}_{y})\hat{\sigma}_{y}=-\hat{\sigma}_{z}\delta/2-\frac{\Omega_{0}}{2}\sin[k_{R}(x-y)]\hat{\sigma}_{y}$, which corresponds to 1D Raman lattice in the $\hat{x}$$-$$\hat{y}$ direction.
We can see that 1D Raman lattice in the $\hat{x}$$-$$\hat{y}$ direction appears in the absorption images for the momentum-resolved Raman spectroscopy as shown in Fig.~\ref{lattice-red}(a2).
When $B_{0}$ is reversed to the $-z$ direction that corresponds to the setting with $(\varphi_{1},\varphi_{2})=(\pi/2, 0)$ as shown in Fig.~\ref{lattice-red}(b1), $\Omega=-i(\bar{M}_{x}+\bar{M}_{y})$ and $\mathbf{\Omega}\cdot \hat{\mathbf{\sigma}}= -\hat{\sigma}_{z}\delta/2-(\bar{M}_{x}+\bar{M}_{y})\hat{\sigma}_{y}=-\hat{\sigma}_{z}\delta/2+\frac{\Omega_{0}}{2}\sin[k_{R}(x+y)]\hat{\sigma}_{y}$, which corresponds to 1D Raman lattice in the $\hat{x}$$+$$\hat{y}$ direction as shown in Fig.~\ref{lattice-red}(b2).
Therefore, this case corresponds to the chiral 2D Raman coupling for SOC, in which two orthogonal 1D Raman lattices appear for the positive and negative directions of the magnetic quantization axis respectively.

When we consider the setting with $(\varphi_{1},\varphi_{2})=(0,0)$ as shown in Fig.~\ref{lattice-red}(c1) and (d1), in which the polarization of the Raman laser $L_{1}$ and $L_{2}$ is linear at the $\pm\pi/4$ angle to the vertical direction respectively, $\Omega=\bar{M}_{x}-i\bar{M}_{y}$ and $\mathbf{\Omega}\cdot \hat{\mathbf{\sigma}}= -\hat{\sigma}_{z}\delta/2+\bar{M}_{x}\hat{\sigma}_{x}-\bar{M}_{y}\hat{\sigma}_{y}$, which corresponds to 2D Raman lattice in the $\hat{x}$$+$$\hat{y}$ and $\hat{x}$$-$$\hat{y}$ direction.
We can see that the 2D Raman lattice in the $\hat{x}$$+$$\hat{y}$ and $\hat{x}$$-$$\hat{y}$ direction appears in the absorption images for the momentum-resolved Raman spectroscopy as shown in Fig.~\ref{lattice-red}(c2).
When $B_{0}$ is reversed to the $-z$ direction, $\Omega=\bar{M}_{x}-i\bar{M}_{y}$ and $\mathbf{\Omega}\cdot \hat{\mathbf{\sigma}}= -\hat{\sigma}_{z}\delta/2+\bar{M}_{x}\hat{\sigma}_{x}-\bar{M}_{y}\hat{\sigma}_{y}$, which corresponds to the similar 2D Raman lattice in the $\hat{x}$$+$$\hat{y}$ and $\hat{x}$$-$$\hat{y}$ direction as shown in Fig.~\ref{lattice-red}(d2).
Therefore, this case corresponds to the non-chiral 2D Raman coupling for SOC, in which the same 2D Raman lattice appears for the positive and negative directions of the magnetic quantization axis respectively.
Here, we would like to mention that this case corresponds 2D SOC based on the optical Raman lattice scheme for studying the topological energy band~\cite{Shuai2018PRL,shuai2019PRL}.

Furthermore, consider the Raman transition $\delta m_{F}=0$ between two hyperfine Zeeman states $|F=7/2, m_{F}=1/2\rangle$$\equiv$$\left\vert \uparrow \right\rangle$ and $|F=9/2, m_{F}=1/2\rangle$$\equiv$$\left\vert \downarrow \right\rangle$.
For this nonchiral transition, only the component of $\mathbf{\hat{e}_{z}}$ of the effective AC magnetic field $\textbf{\emph{B}}_{AC}^{eff}$ (Eq.~(\ref{HH24})) parallel to the static bias field $\textbf{\emph{B}}_{0}$ can drive transition $\delta m_{F}=0$ and the components of $\mathbf{\hat{e}_{x}}$ and $\mathbf{\hat{e}_{y}}$ perpendicular to the static bias field $\textbf{\emph{B}}_{0}$ are neglected.
We can see that the same 2D Raman lattice in the $\hat{x}$$+$$\hat{y}$ and $\hat{x}$$-$$\hat{y}$ direction appears in the absorption images for the momentum-resolved Raman spectroscopy in the positive and negative directions of the magnetic quantization axis respectively, as shown in Fig.~\ref{different-transition}(a).
Therefore, this case corresponds to the non-chiral 2D Raman coupling for SOC.
Theoretically, the atomic absorption images from Fig.~\ref{lattice-red}(c2) and (d2), as well as Fig.~\ref{different-transition}(b2) and (c2) should display exactly the same pattern. However, there are many factors to cause some residual difference of the atomic response in the realistic experiment. For example, when the magnetic field flips from the +z direction to the -z direction, it is not possible to ensure a perfect 180$^{\circ}$ flip, which leads to a deviation of the effective AC magnetic field relative to the bias magnetic field. Moreover, the fluctuation of atom number and Raman laser intensity can also cause the difference of the momentum-resolved Raman spectroscopy.

Finally, consider the Raman transition $\delta m_{F}$=$0,\pm1$ based on two same polarization in x-y plane of Raman lasers as shown in Fig.~\ref{paipai}(a), which drive the Raman transition $\delta m_{F}$=$0$ ($|9/2, 1/2\rangle$$\rightarrow$$|7/2, 1/2\rangle$) and $\delta m_{F}$=$\pm1$ ($|9/2, 1/2\rangle$$\rightarrow$$|7/2, -1/2\rangle$), respectively.
For this nonchiral Raman transition $\delta m_{F}$=$0$ as shown in Fig.~\ref{paipai}(b), the component of $\mathbf{\hat{e}_{z}}$ for the effective AC magnetic field $\textbf{\emph{B}}_{AC}^{eff}$ (Eq.~(\ref{HH24})) parallel to the static bias field $\textbf{\emph{B}}_{0}$, which is same to the case in  Fig.~\ref{different-transition}(a).
However, for the transition $\delta m_{F}$=$\pm1$, there is no Raman coupling as shown in Fig.~\ref{paipai}(c), since the only component of $\mathbf{\hat{e}_{z}}$ for AC magnetic field $\textbf{\emph{B}}_{AC}^{eff}$ parallel to the static bias field $\textbf{\emph{B}}_{0}$ can not drive transition $\delta m_{F}=\pm1$ to induce disappearance of Raman coupling.

In conclusion, we present the intuitive picture of the chirality of spin motion in a static and AC magnetic field by the Bloch sphere to better understand the physics of the non-chiral and chiral Raman coupling for SOC.
We study the chiral Raman coupling created by the chiral light-atom interaction, in which a circularly polarized AC magnetic field generated by two Raman lasers interacts with the chiral transition $\delta m_{F}$=$\pm 1$ between two Zeeman spin states.
The chirality of Raman coupling for SOC is demonstrated by setting different laser polarization configurations under the different two-photon transition rules. 1D and 2D chiral Raman coupling for SOC are investigated experimentally in detail.
This work provides a new system for understanding chiral physics and investigating the states of matter with chirality.

\begin{acknowledgments}
This research is supported by National Key Research and Development Program of China (Grants No. 2022YFA1404101, and No. 2021YFA1401700), Innovation Program for Quantum Science and Technology (Grants  No. 2021ZD0302003), National Natural Science Foundation of China (Grants No. 12034011, No. U23A6004, No. 12474266, No. 12474252, No. 12374245, No. 12322409, and No. 92065108), and the Fund for Shanxi 1331 Project Key Subjects Construction.
\end{acknowledgments}

\bibliography{references-Bloch}

\end{document}